\documentclass[aps,preprintnumbers,pre,showpacs,superscriptaddress]{revtex4}
\usepackage{amssymb}
\usepackage{amsmath}
\usepackage{amsopn}
\usepackage{graphics}
\usepackage{graphicx}
\usepackage{epsfig}

\begin{document}

\title{Study of the interplay between magnetic shear and resonances using Hamiltonian models for the magnetic field lines}
\author{M.-C. Firpo}
\email{marie-christine.firpo@lpp.polytechnique.fr}
\affiliation{Laboratoire de Physique des Plasmas, CNRS - Ecole
Polytechnique, 91128 Palaiseau cedex, France}
\author{D. Constantinescu}
\affiliation{Dept of Applied Mathematics, Association Euratom-MECI,
University of Craiova, Craiova 200585, Romania}

\begin{abstract}
The issue of magnetic confinement in magnetic fusion devices is
addressed within a purely magnetic approach. Using some Hamiltonian
models for the magnetic field lines, the dual impact of low magnetic
shear is shown in a unified way. Away from resonances, it induces a
drastic enhancement of magnetic confinement that favors robust
internal transport barriers (ITBs) and stochastic transport reduction. When
low-shear occurs for values of the winding of the magnetic field
lines close to low-order rationals, the amplitude thresholds of the
resonant modes that break internal transport barriers by allowing a
radial stochastic transport of the magnetic field lines may be quite
low. The approach can be applied to assess the robustness versus
magnetic perturbations of general (almost) integrable magnetic
steady states, including non-axisymmetric ones such as the important
single helicity steady states. This analysis puts a constraint on
the tolerable mode amplitudes compatible with ITBs and may be
proposed as a possible explanation of diverse experimental and
numerical signatures of their collapses.
\end{abstract}


\maketitle

\section{Introduction}

\label{sec_intro} In the present hot plasma toroidal devices, it is widely
recognized that the essential limitation that prevents ignition to be
reached is due to an insufficient energy confinement resulting from
small-scale turbulence. During the last two decades, some decisive progress
has however been made to increase the energy confinement time through the
widespread experimental achievement of internal transport barriers (ITBs)
\cite{WolfRevue,ConnorReviewITB2004}. In these regimes, there exists a
region of the tokamak plasma core where the local ion and electron thermal
diffusivities are substantially reduced, nearly down to their neoclassical
values \cite{ConnorReviewITB2004}. This is likely to be a signature of the
regularity of the central magnetic surfaces. In this way, the emergence of
ITBs results in a sort of virtuous circle for fusion with both high central
temperature $T$ and high energy confinement time $\tau_{E}$ increasing the
triple product $n_{e} T \tau_{E}$, which implies a better fusion yield. The
importance of the magnetic shear, through the form of safety factor
profiles, in reaching these states by quenching the core plasma turbulent
transport has been firmly established. Decisive ingredients were given
notably by the experimental results obtained by Eriksson et al. \cite%
{Eriksson02} who demonstrated experimentally that the $q$-profile could be
used as a single control parameter to obtain ITBs. In these experiments,
they could obtain ITBs for non-monotonic $q$-profiles but not with
monotonously growing $q$, while other parameters such as the $\mathbf{E}%
\times \mathbf{B}$ flow shear were kept the same. Additional evidence was
later given by Sauter et al. \cite{Sauter2005} through a dedicated
experiment. They controlled the $q$-profile using inductive current to
generate positive and negative current density perturbations in the plasma
center, with negligible input power. In this way, they were able to
demonstrate that the electron confinement could be modified significantly
solely by perturbing the current density profile. Other experiments in
various tokamaks showed that having an inner region with low magnetic shear
was indeed sufficient to obtain ITBs \cite{Razumova2001,ConnorReviewITB2004}.

In reverse field pinches (RFPs) also, improved performance related
to magnetic chaos healing \cite{Escande2000} has been recently
achieved experimentally, yet through non-axisymmetric states
\cite{LorenziniNatureP,Puiatti2009,Bonfiglio2010}. It was observed
that, by increasing the plasma current, the RFP operated some change
of magnetic topology through a spontaneous transition towards a
single-helicity equilibrium strongly resilient to magnetic chaos.

These various experimental results emphasize the importance of the
magnetic structure as they show that some configurations -
reversed-shear, or low-shear, $q$-profiles in tokamaks or some
single-helicity equilibria in RFPs - enable a substantial
improvement of the confinement. This is an invitation to consider a
purely magnetic approach of the problem \cite{Nasi2009} that will be
pursued here. This amounts to address the issue of the plasma
confinement in terms of the magnetic confinement, namely in terms of
the confinement of the magnetic field lines. The ensuing approach is
thus obviously a simplified approximation of a more complex,
self-consistent, reality involving a larger number of effective
degrees of freedom, including in particular the electric field. It
is a sort of zero order approach that relies on the essential
principle of magnetic confinement devices which is to use the
magnetic field to confine the charged particles composing the fusion
fuel in a plasma state \cite{Morrison2000}. In this sense, ensuring
the confinement of the magnetic field lines may be viewed as a
prerequisite to particle and energy confinement. As it is well
known, this purely magnetic approach may be addressed within an
Hamiltonian formalism that results from the universal
divergence-free nature of $\mathbf{B}$ \cite{Boozer2004}. An
intermediate, first-order, step towards a fully realistic
description of the physics of magnetic confinement devices would be
to treat the test particle motion \cite{AbdullaevPoP2006} instead of
the magnetic field lines. This is left for a future work.

In the absence of resonances, that will be specified below, the
dramatic improvement on magnetic confinement induced by a low
magnetic shear condition will be recalled in Section
\ref{sec_benefits} and its link to ITBs and core stochastic transport
reduction discussed. The improved confinement observed in RFPs in
single-helicity states will be shown to result from the same
mechanism when moving to appropriate action-angle variables. However
this optimistic scenario will be tempered since, in actual tokamaks,
ITBs may eventually collapse. In Section \ref{sec_drawbacks}, it
will be shown, using some Hamiltonian models for the magnetic field
lines, that this low shear condition happens to be strongly
deleterious to magnetic confinement when approaching resonances by
enabling stochasticity for much lower magnetic perturbation
amplitudes. This will be further analyzed in Section
\ref{sec_analysis}. A discussion of this phenomenon and of its
possible consequences will be finally given in Section
\ref{sec_conclusion}.

\section{Benefits of a low magnetic shear}

\label{sec_benefits}

\subsection{Hamiltonian formulation for the magnetic field lines}

It is well known that the equations of magnetic field lines can be
cast in Hamiltonian form. Actually, the field lines of a magnetic,
$\mathbf{B}(\mathbf{x})$, or any other divergence-free field are the
trajectories of a Hamiltonian (See e.g. the review by Boozer
\cite{Boozer2004} or Abdullaev's book
\cite{AbdullaevContructionMappings2006}).

In tokamaks, a dependance on the toroidal angle breaking the
rotational invariance is analogous to the usual time-dependence in
Hamiltonian dynamical systems. Generically one is thus left with a
one-and-a-half degrees of freedom Hamiltonian, that can be
decomposed into an axisymmetric, integrable part and a perturbation
term that may be Fourier decomposed in the poloidal and toroidal
angles. For tokamaks, the Hamiltonian reads
\begin{equation}
H(\psi,\theta,\phi)=H_0(\psi)+\sum_{m,n}H_{mn}(\psi)\cos(m \theta - n
\phi+\chi_{mn}),  \label{hamil}
\end{equation}
with $H_0(\psi)=\int^{\psi}W(\psi^{\prime})d \psi^{\prime}$ and $%
W(\psi)=1/q(\psi)$ the winding profile, inverse of the safety factor
profile. In the tokamak configuration, $H$ is the poloidal magnetic flux and
$\psi$ the toroidal magnetic flux conjugated to the poloidal angle $\theta$.

\subsection{Some results on the benefits of low shear}

\label{sub_tokamap} In real magnetic fusion devices, the magnetic
perturbations given by the wave amplitudes $H_{mn}(\psi)$ may be
induced by a variety of phenomena. They may result from external
effects, such as the ripple, or may result from intrinsic phenomena
such as MHD activity. In order to address the magnetic confinement
problem, one may turn to simplified models for the form of the
perturbed magnetic field. Restricting to the single poloidal number
$m=1$, taking equal phases $\chi_{1n}$ and some special form of the
radial perturbation $H_{1n}(\psi)$ identical for all $n$'s yields
the continuous version of the tokamap \cite{BalescuTokamap1998}.
Then, considering the limit of a large number of toroidal modes
makes the associated Poincar\'{e} (toka)map appear naturally through
the use of periodic $\delta$ distributions.

Using the symmetric tokamap framework, it was recently shown in Ref.
\cite{Nasi2009} that the existence of a low magnetic shear region
produced a drastic enhancement of the confinement of the magnetic
field lines. More
specifically, the existence of some $\psi_s$ such that $H_0^{\prime\prime}(%
\psi_s)=0$, that amounts to $W^{\prime}(\psi_s)=0$, was shown to
improve significantly the robustness w.r.t. magnetic perturbations
of the regular core magnetic field lines inside the $\psi=\psi_s$
surface, acting then as a transport barrier, in agreement with
previous results obtained in so-called non-KAM systems including
mainly maps with reversed shear profiles (See e.g.
\cite{dCN1996,PhysRevE.58.3781,Constantinescu2005,Wingen2005,Portela2008,RypinaRobustTransportBarriers2007}).
Interestingly, Ref. \cite{Nasi2009} gave evidence that this effect
does not necessitate $q$ to be reversed shear: the $q$-profile can
also be monotonous with an inflexion point, which agrees with the
diversity of ITBs experimental scenarios. The beneficial effect of
the radial extent of the low shear zone for safety profiles with
$q>1$ was demonstrated. Recently, these results were more rigorously
grounded: Theorems of existence of invariant circles
for non-twist maps were formulated for the symmetric tokamap \cite%
{ConstanFirpo} using a theorem about small shear systems formulated
by Ortega \cite{Ortega}.

\subsection{Extension to non-axisymmetric states}

It is important to note that the domain of applicability of this favorable
regime may be extended from axisymmetric equilibria to non-axisymmetric
integrable states provided $\psi$ is replaced by some suitable action
variable. Let us consider the case where some magnetic configuration
involves not only an axisymmetric component $H_0(\psi)$ but also some $%
(m_0,n_0)$ component whose amplitude $A$ is much larger than that of the
other $(m,n)$ waves of order $\varepsilon$. The Hamiltonian reads then
\begin{equation}
H(\psi,\theta,\phi)=H_0(\psi)+A H_{m_0 n_0}(\psi)\cos(m_0 \theta -n_0 \phi),
\label{HamiltonianAepsilon}
\end{equation}
up to $\mathcal{O}(\varepsilon)$ terms. The dominant part of the
Hamiltonian, given in Eq. (\ref{HamiltonianAepsilon}), may be put into an
explicitly integrable form
\begin{equation}
\overline{H}(\Psi,\Theta)=H_0(m_0 \Psi)-n_0 \Psi+A H_{m_0 n_0}(m_0 \Psi)\cos
\Theta
\end{equation}
through a canonical change of variables, with the generating function $%
F_2(\Psi,\theta,\phi)=(m_0 \theta - n_0 \phi) \Psi$, that amounts to moving
to the $(m_0,n_0)$ wave frame. Even if $H_0^{\prime\prime}(\psi)$ does not
vanish, it is then easy to recognize that, provided $A$ is large enough, a
shearless condition in the action variable
\begin{equation}
J\equiv \frac{1}{2 \pi} \oint \Psi d\Theta
\end{equation}
associated to the integrable part of the Hamiltonian
(\ref{HamiltonianAepsilon}) may however be obtained. In this way,
there can be non-axisymmetric magnetic states that are much more
resilient to magnetic chaos than what would be their axisymmetric
components. An illustration of this phenomenon was given in Ref.
\cite{Escande2000}. It was shown that the magnetic structure
involving some small $m=1$ magnetic island with a separatrix was
less robust towards magnetic perturbations than a reconnected
structure with a higher $m=1$ amplitude. This transition in terms of
the winding profile in the action variable $J$ may be interpreted as
follows: In the low amplitude case where a magnetic island is
present, the separatrix is associated to a vanishing of $W(J)$
within a cusp singularity, whereas, as the amplitude grows, the
reconnection process eventually makes the separatrix disappears
inducing locally the growth of $W(J)$ and smoothing the singularity,
previously associated to the separatrix, which causes the appearance
of a local minimum of $W(J)$. This later favorable process may be
responsible of the unexpectedly good confinement of the
non-axisymmetric self-organized states very recently achieved in a
RFP \cite{LorenziniNatureP} and in a compact toroidal
\cite{CothranPRL2009} plasmas. For the tokamak setting, this
non-axisymmetric formulation of the low shear or zero-twist
condition may open new perspectives to improve magnetic confinement.

\section{Detrimental effect of low shear in the vicinity of resonances using some Hamiltonian models}

\label{sec_drawbacks}

Let us now go back to the notations introduced in Eq. (\ref{hamil}). The
above analysis did not take into account the possibility of resonances. In
particular, the fact that the winding number $W_s \equiv W(\psi_s)$,
associated to the action $\psi_{s}$ such that $H_0^{\prime\prime}(\psi_s)=0$%
, could be in the vicinity of some phase velocity $n/m$ corresponding to the
$(m,n)$ magnetic perturbation was not considered. For instance, works using
the tokamap framework \cite{BalescuTokamap1998,nobleInternal,Nasi2009}
ignored de facto this possibility by restricting to the $m=1$ case while
taking typically $q$-profiles above 1 (with the noticeable exception of the
sawtooth-related studies of Ref. \cite{Pavlenko}).

In order to illustrate the interplay between low shear and resonances, let
us first consider a simplification of the system (\ref{hamil}) similar to
the continuous tokamap case but where only the $m=4$ poloidal modes are
retained, with
\begin{equation}
H\left( \psi ,\theta ,\phi \right) =\int\limits^{\psi }\frac{dx}{q(x)} +K
\widehat{H}(\psi ) \sum_{n=-M}^{M}\cos \left(4\theta -n\phi\right).
\label{h_continuous_mequal4}
\end{equation}
For $M \geq 1$, this represents a one-and-a-half degrees of freedom
Hamiltonian in which the perturbation amounts to a superposition of
plane waves with winding numbers equal to $n/4$. It is convenient,
and physically relevant, to consider the large $M$ case for which
the derivation of the symplectic mapping associated to
(\ref{h_continuous_mequal4}) is immediate through the use of the
periodic-$\delta$ distribution. For computational purposes, we used
a symmetric version
\cite{AbdullaevContructionMappings2006,Wingen2005} of the later
mapping that yields a very accurate modeling of the continuous
dynamics (\ref{h_continuous_mequal4}) with
$\widehat{H}(\psi)=2\psi/(1+\psi)$. The physical motivation to
consider this specific form of magnetic perturbation was discussed
in Ref. \cite{BalescuTokamap1998}. Basically, this form roughly
mimics the radial (or $\psi$) form of MHD tearing modes. It gives a
vanishing perturbation in the limit $\psi \rightarrow 0$ and ensures
that the polar axis cannot be crossed so that, under the map, $\psi$
remains positive as it should be. It would have been possible to use
fully realistic MHD profiles for each $(m,n)$ mode obtained from
experiments (as e.g. in the study of Ref. \cite{Igochine2006}). Our
point in the present Section is however to use a simpler framework
in order to emphasize the specific role of $W_{s}$.

In order to use $W_s$ as a single control parameter, we considered
the family of winding profiles $W_p(\psi)=W_s-c(\psi-\psi_s)^2$ with
fixed $c$ and $\psi_s$. These correspond to reversed shear
$q$-profiles with zero
shear at $\psi_s$. To be specific, $W_s$ was allowed to vary while putting $%
c=.5$ and $\psi_s=0.4$ in the whole study.
\begin{figure}[tbph]
\begin{center}
\includegraphics[width = .7 \textwidth]{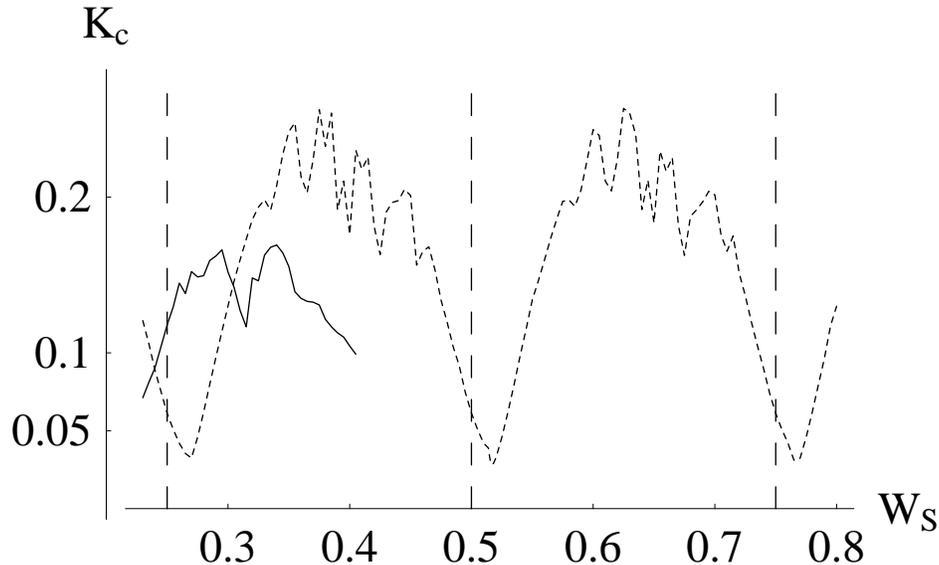}
\end{center}
\caption{The critical value of the stochasticity parameter $K$, as defined
in the text, as a function of the inverse of the minimum of the $q$-profile,
$W_s$, for the parabolic family $W_p(\protect\psi)$ (dashed line) and as a
function of the parameter $W_s$ in the linear case $W_{l}(\protect\psi)$
(plain line). The vertical dashed lines mark the locations of the resonances
$W_s=1/4$, $2/4$ and $3/4$.}
\label{plot_Kc}
\end{figure}

In order to quantify the impact of the specific value of the minimum of the $%
q$-profile, $q_{\mathrm{min}}$, (or of its inverse $W_{s}=1/q_{\mathrm{min}}$%
), we defined by $K_c$ the value of the stochasticity parameter $K$ above
which there exists some initial condition with $\psi<0.1$ whose evolution
under (\ref{h_continuous_mequal4}) eventually escapes the physically
accessible phase space by crossing $\psi=1$. Practically, for each value of $%
W_s$ and a given value of $K$, up to one hundred initial conditions
were chosen at random in the range $0<\psi<0.1$ and allowed to
evolve under the symmetric map obtained from Eq.
(\ref{h_continuous_mequal4}) for up to 5000 iterations. If some
$\psi$ happened to cross $\psi=1$, the process was relaunched for a
smaller $K$ or else, for a larger $K$, in a dichotomic way, so as to
converge towards the percolation-like threshold $K_c$ value. The
obtained results have been plotted in Fig. \ref{plot_Kc}. Let us
first briefly comment on the meaning of the indicator $K_{c}$. Using
the form of the magnetic perturbations as
$\widehat{H}(\psi)=2\psi/(1+\psi)$ ensures that core magnetic field
lines remain regular whereas border field lines (at $\psi=1$) are
the first to feel chaos as $K$ increases. Having in mind fusion
applications, $K_{c}$ has a practical meaning in terms of a measure
of the confinement of the magnetic field lines, as it gives a
threshold above which magnetic field lines starting from the core
eventually reach the border. It is also an indicator of the global
quantitative measure of the stochasticity of the magnetic field
lines since, in the present model, there cannot be a connection
between the core and the border of the $\psi$-domain through a
regular magnetic field trajectory. However, with respect to the
magnetic confinement issue, $K_{c}$ may underestimate chaos, in the
sense that there may exist non-negligible subsets of the physical
central phase space domain in which chaotic field lines are chaotic
already for values of $K$ below $K_{c}$.

\begin{figure}[tbp]
\begin{center}
\includegraphics[width =  .7 \textwidth]{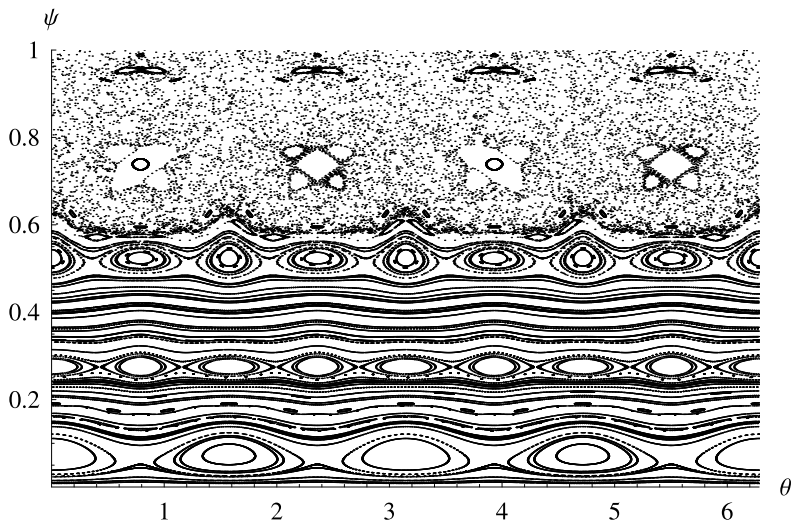}
\end{center}
\caption{Poincar\'{e} plot of the system (\protect\ref{h_continuous_mequal4}%
) with the parabolic reversed-shear profile $W_p(\protect\psi)$ for $K=0.03$
and $W_s=0.645$.}
\label{Poincaré0p645}
\end{figure}

It is straightforward to realize that, for the profiles $W_p(\psi)$ or $%
W_l(\psi)\equiv W_s-c(\psi-\psi_s)$, the dynamics is invariant under a
translation in $W_s$ of some multiple of $1/4$. Effectively the curve giving
$K_c$ as a function of $W_s$ is $1/4$-periodic in $W_s$ up to statistical
discrepancies. Some part of the curve obtained for the linear strong shear
case $W_l(\psi)$ has been drawn for comparison for about one period in $W_s$%
. Basically, the effect of low shear happens to be efficient for $W_s$
between the $n/4$ resonances by increasing significantly $K_c$ w.r.t. the
high shear case. The Poincar\'{e} plot of Figure \ref{Poincaré0p645}
corresponds to such a favorable situation. This reflects the fact that, in
spite of a non-negligible perturbation of magnitude $K$, winding profiles
having a low shear region can sustain almost regular core magnetic surfaces.
This magnetic chaos healing is likely to explain the stochastic transport reduction in
ITBs due to the \emph{sole} effect of the $q$-profile reported in Refs. \cite%
{Eriksson02,Sauter2005}.

However, in the vicinity of resonances, this tendency reverses: low
shear induces abrupt drops of the magnetic confinement whereas quite
higher threshold values of $K$ are reached for the high-shear linear
winding profile. In this case, the statement formulated by
Rosenbluth, Sagdeev, Taylor and Zaslavski \cite{Rosenbluth1966} that
strong shear is beneficial to preserve the regularity of the
magnetic surfaces applies.

\begin{figure}[htbp]
\begin{center}
\includegraphics[width =  .7 \textwidth]{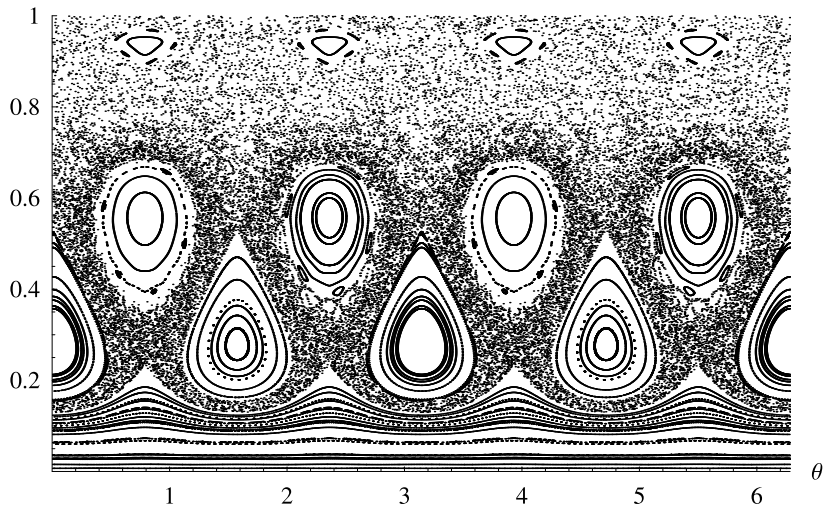}
\end{center}
\caption{Poincar\'{e} plot of the system (\protect\ref{h_continuous_mequal4}%
) with the parabolic reversed-shear profile $W_p(\protect\psi)$ for $K=0.03$
and $W_s=0.52$.}
\label{Poincaré0p52}
\end{figure}

For the reversed-shear winding profile $W_{p}(\psi)$, low minima of
$K_c$ occur for values of $W_s$ slightly above the linear values of
the resonances given by $n/4$. Figure \ref{Poincaré0p52} depicts
what happens in this case. The abrupt degradation of magnetic
confinement in the vicinity of the resonances coincides with the
emergence of some local stochastic web  pattern \cite{MinimalChaos,WeakChaos,note} that
enables a large diffusion in the action $\psi$. Under the
perturbation, separatrices of the island chains, a characteristic
phase-space pattern of the non-twist Hamiltonian systems, are
destroyed and replaced with stochastic channels of finite width,
visible on this Figure. This produces a path to some deleterious
radial transport already at low values of the amplitude of the
resonant mode. The merging of separatrices is the phenomenon through
which global chaos enters the system and will be more closely
examined in Section \ref{sec_analysis}. In another physical context,
Soskin \textit{et al.} studied a close phenomenon of nonlinear
resonance in the minimal case of a single exciting wave
\cite{Soskin}. Here, the consideration of a large mode number $M$,
that is physically relevant to account for turbulence in toroidal
fusion devices, should provide an additional effective randomization
of motion in the vicinity of the separatrix.

\begin{figure}[tbph]
\begin{center}
{\includegraphics[width = .7 \textwidth]{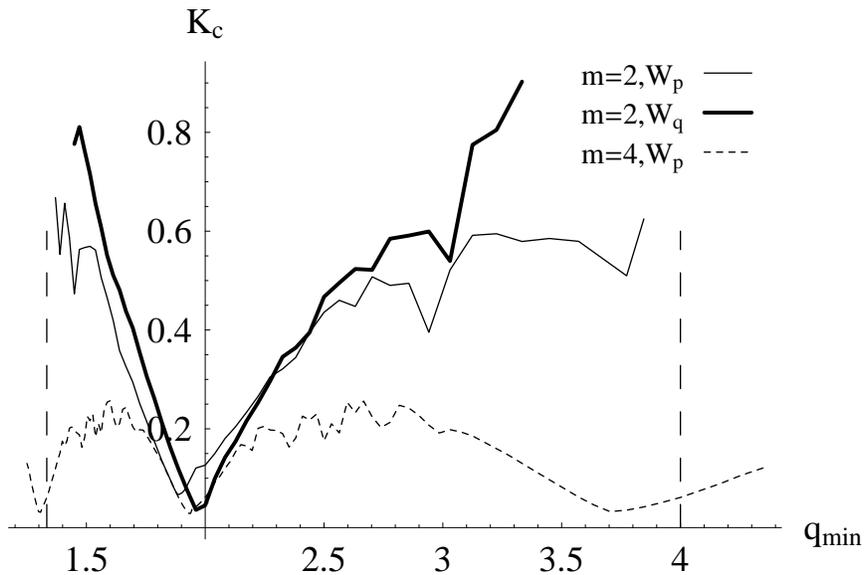}}
\end{center}
\caption{Critical value of the stochasticity parameter $K$, as defined in
the text, as a function of the minimum of the $q$-profile, $q_{\mathrm{min}%
}\equiv 1/W_s$.}
\label{KcvsQmin}
\end{figure}

All other things being equal, we then considered the Hamiltonian system for
the magnetic field lines (\ref{h_continuous_mequal4}) obtained by just
replacing $m=4$ by $m=2$ and computed its associated symmetric map. Figure %
\ref{KcvsQmin} displays the obtained values of the threshold $K_c$ defined
for two cases of winding profiles, the parabolic one $W_p(\psi)$ and the
quadratic one $W_q(\psi) \equiv W_s-c(\psi-\psi_s)^4$, together with the
curve previously obtained for $m=4$ for the reversed-shear parabolic
profile. This time $K_c$ is plotted as a function of the minimum of the
safety factor profile $q_{\mathrm{min}}\equiv 1/W_s$, a parameter more
commonly used by the tokamak community.
In the $m=2$ case, the positive role on magnetic confinement induced by the
existence of a low shear region away from resonances appears in full
clarity. The threshold $K_c$ attains there very large values and the
comparison with the $m=4$ case is illuminating: the span in $W_s$ between
two subsequent $m=2$ resonances is twice larger than the one between two
subsequent $m=4$ resonances and $K_c$ is then allowed to grow to larger
amplitudes between resonances. The effect of the amount of flatness around $%
q_{\mathrm{min}}$ is investigated by comparing the results obtained for $W_p$
to that for the quadratic flattest profile $W_q$ in the $m=2$ case. Away
from resonances, it is clear that expanding the low shear zone improves the
resilience of the magnetic structure towards perturbations as shown in Ref.
\cite{Nasi2009}. However, in the vicinity of resonances, the effect
reverses. A lower minima of $K_c$ is obtained for $W_q$ and it is reached
closer to the resonance $q=2/n$ (with $n=1$ in Fig. \ref{KcvsQmin}).

\section{Local analysis}

\label{sec_analysis}

We wish to interpret the above results within the continuous Hamiltonian
system. Let us consider then the Hamiltonian%
\begin{equation}
H\left( \psi ,\theta ,\phi \right) =\int\limits^{\psi }W(x)dx+K\widehat{H}%
(\psi )\sum_{n=-M}^{M}\cos \left( m_{0}\theta -n\phi \right)
\end{equation}
for some given poloidal mode $m_{0}$ and take $W(\psi )=W_{s}-c(\psi -\psi
_{s})^{k}$, with $c>0$ and $W_{s}=W(\psi _{s})$. Let us proceed to a
canonical change of variable with the generating function%
\begin{equation}
F_{2}(\psi ^{\prime },\theta ,\phi )=\left( m_{0}\theta -n_{0}\phi \right)
\psi ^{\prime }+\psi _{s}\theta .
\end{equation}
This yields%
\begin{eqnarray}
\frac{\partial F_{2}}{\partial \theta } &=&\psi =m_{0}\psi ^{\prime }+\psi
_{s}, \\
\frac{\partial F_{2}}{\partial \psi ^{\prime }} &=&\theta ^{\prime
}=m_{0}\theta -n_{0}\phi ,
\end{eqnarray}
and the new Hamiltonian reads, up to a constant,
\begin{equation}
H\left( \psi ^{\prime },\theta ^{\prime },\phi \right) =\left(
m_{0}W_{s}-n_{0}\right) \psi ^{\prime }-\frac{c}{k+1}(m_{0}\psi
^{\prime })^{k+1}+K\widehat{H}\left( m_{0}\psi ^{\prime }+\psi
_{s}\right) \sum_{n=-M}^{M}\cos \left( \theta ^{\prime }+n_{0}\phi
-n\phi \right) .
\end{equation}
Sufficiently close to the resonance $n=n_{0}$, $\theta ^{\prime }+n_{0}\phi
-n\phi $ is slowly varying, so that one can average over the fast "time" $%
\phi$. This yields the one degree of freedom Hamiltonian%
\begin{equation}
H\left( \psi ^{\prime },\theta ^{\prime },\phi \right) =\left(
m_{0}W_{s}-n_{0}\right) \psi ^{\prime }-\frac{c}{k+1}(m_{0}\psi
^{\prime })^{k+1}+K\widehat{H}\left( m_{0}\psi ^{\prime }+\psi
_{s}\right) \cos \theta ^{\prime }.  \label{Ham_onereson}
\end{equation}
The study of its topology as a function of $W_{s}$ and $K$ should help to
clarify the phase space location where the stochasticity, induced by the
perturbation to (\ref{Ham_onereson}), should first emerge as $K$ increases
from zero.
\begin{figure}[tbph]
\begin{center}
{\includegraphics[width = \textwidth]{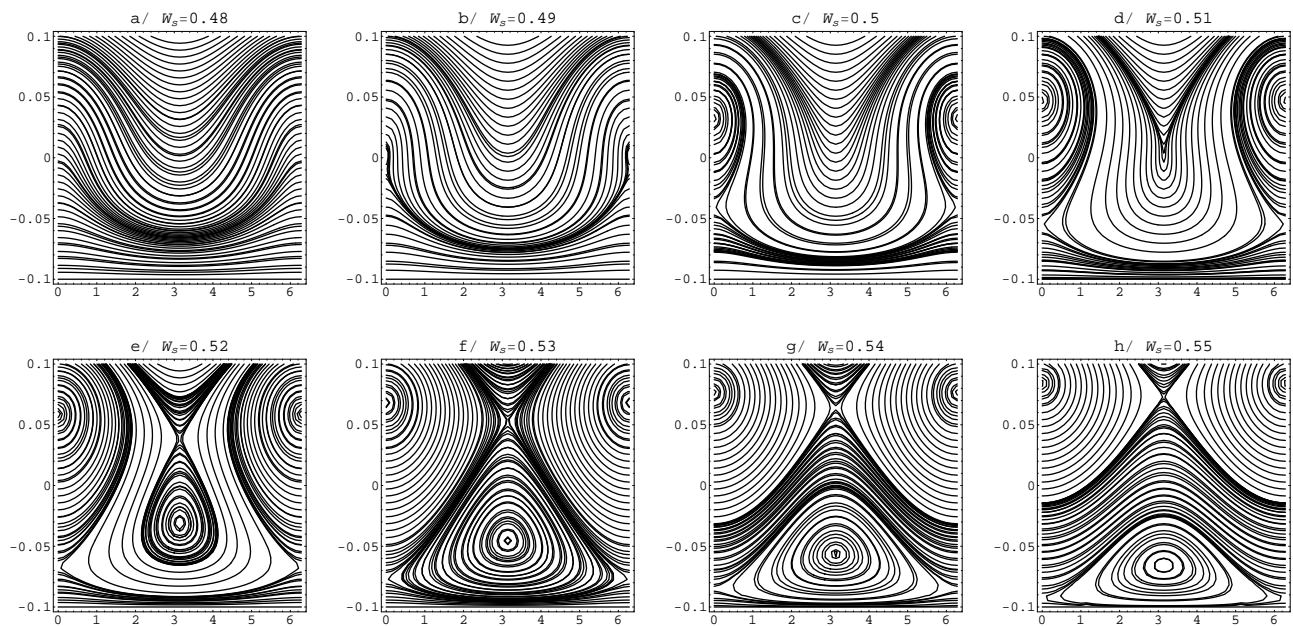}}
\end{center}
\caption{Phase space of the Hamiltonian (\ref{Ham_onereson}) for
some values of $W_{s}$ close to $n_{0}/m_{0}$ with $m_{0}=4$ and
$n_{0}=2$, for $c=0.5$ and $K=0.01$.} \label{ContourPlot}
\end{figure}
Figure \ref{ContourPlot} shows the transition from an homoclinic
topology with dimerized islands (plot e) to an heteroclinic topology
with twin islands (plots g and h) as $W_s$ increases while being
slightly above $n_{0}/m_{0}=1/2$. The bifurcation between these two
topologies \cite{VDWeele} occurs at some reconnection threshold
$W_{s}^{*}$ (see e.g. Ref. \cite{Howard1984,DDCNMCF}) that depends
on the perturbation amplitude $K$ and can be easily computed by
equaling the energy values at the x-points obtained for $\theta=0$
and $\theta=\pi$.

As it is well known, the nonintegrable perturbation coming from the
other modes will gradually replace, as $K$ increases, separatrices
by stochastic layers. On the basis of Figure \ref{ContourPlot}, one
can argue that the largest extent of the stochastic domain shall
take place in the case where $W_s$ is slightly above $n_{0}/m_{0}$,
close to the situation corresponding to the plot e, in which case
the stochastic layers coming from the two separatrices shall merge
for a sufficiently large value of the perturbation amplitude. In
this way, this analysis gives account of the fact that the smallest
values of the critical thresholds take place, for the given
$\psi$-form of the perturbation, not for $W_{s}=n_{0}/m_{0}$ but
slightly larger values. The Reader is referred to
\cite{AbdullaevContructionMappings2006,scholarpedia} for extra
reference studies on the transition to chaos in Hamiltonian systems
with non-monotonic frequency.

\section{Discussion}

\label{sec_conclusion}

The previous qualitative, yet systematic, study has emphasized the
dual impact of low magnetic shear on magnetic confinement in
toroidal devices. The main achievement of this approach has been to
hopefully clarify the dual impact of low magnetic shear through a
unified picture. In particular, it has been shown using some
Hamiltonian models for the magnetic field lines that, away from any
consideration on dynamics and stability, there exists potentially
very low thresholds on magnetic perturbations above
which safety factor profiles having a low shear region occurring for $q$%
-values close to $q=m/n$ induce a loss of magnetic confinement. This
may now be used both to predict and to interpret experimental
observations.

For instance, assume that some reversed-shear $q$-profile associated
to an ITB regime evolves in such a way that $q_{\mathrm{min}}$
decreases and crosses e.g. $q=2$. Then, if there exists some
magnetohydrodynamic (MHD) mode with $m=2n$, typically a (2,1) mode,
having a sufficiently large amplitude, one predicts a collapse of
this ITB due to an abrupt macroscopic degradation of magnetic
confinement. This prediction does not need any consideration on the
nature of the instability that produced this particular (2,1) mode.
However, to make a quantitative statement on the amplitude threshold
inducing the collapse one should have some knowledge on the radial
form of the modes involved .

This prediction does not either preclude the fact that
ITBs may be more easy to trigger experimentally for rational values of $q_{%
\mathrm{min}}$ for low MHD activity, a fact on which there does not
seem to be a general agreement \cite{Sauter2005} anyway. There is
some subtlety here: For instance, for the original symmetric
tokamap, that corresponds to the case where only $m=1$
(non-resonant) perturbations are retained, the most robust magnetic
surfaces would actually be obtained for $q_{\mathrm{min}}$ close to
2. It could however explain the fact that the good confinement phase
obtained in these cases is always transient \cite{Tuccillo2006} as
various $(m,n)$ modes should eventually be destabilized either
linearly or nonlinearly through mode couplings. Moreover, as it is
well known from the tokamak linear MHD theory, the low $m$ and $n$
MHD modes are typically the most dangerous ones, in the sense that
they are associated to the largest linear growth rates.
Consequently, one expects drops of magnetic confinement to take
place in the first place for low-order rational values of
$q_{\mathrm{min}}$, as the amplitude thresholds requested for the
large-scale degradation of the magnetic confinement should be first
attained for those low $(m,n)$ modes.

As illustrated in Fig. \ref{KcvsQmin}, the negative impact of low
shear should be either benign and transient in the case where the
$q$-profile happens to rapidly cross low rational values of
$q_{\mathrm{min}}$, or particularly deleterious, if the $q$-profile
happens to be clamped and very flat about $q_{\mathrm{min}}$ close
to some $\overline{q}\equiv m/n$, with a large enough amplitude in
the $(\overline{q}n,n)$ modes. These ITBs collapses triggered by
stochastization of the magnetic field lines should manifest through
flattenings of the pressure and temperature profiles in the
stochastic region.

Interestingly enough, these predictions agree with numerous experimental and
numerical observations on ITBs dynamics. We should first note that
experimental evidences indicate that the so-called Edge-Localized modes
(ELMs) may not be responsible for ITBs collapses since these are observed to
happen already before the ELMs occur \cite{0029-5515-41-10-306}.
Additionally, ITB do collapse while the $q$-profile is reversed-shear and
there is a variety of experimental report of core confinement collapses in
reversed-shear cases when $q_{\mathrm{min}}$ reaches values close to low
order rational values such as 3/2, 2 or 3 \cite%
{JET2002,Takechi2005,0029-5515-41-10-306}. The extensive
experimental and numerical results of Ref. \cite{maget2007}
consistently show that pressure crashes associated to core magnetic
stochasticity are strongly correlated with the closeness of
$q_{\mathrm{min}}$ to 2 and with a sufficient amplitude in the
$(2,1)$ mode. The proposed picture is also consistent with recent
numerical results showing that an ITB collapse occurs for $q_{\min}$
in the vicinity of $4/3$ when turbulence involves a sufficiently
large amplitude of the $(4,3)$ mode \cite{Tokunaga2009}. Moreover,
the above predictions should be most easily evidenced on electron
ITBs, since due to their small Larmor radius, electrons provide a
scan of the magnetic field structure. In this respect, it would be
interesting to investigate further the connections with the
experimental results of Ref. \cite{baar1999} in which the
disappearance of electron transport barriers are associated with
$q_{\mathrm{min}}$ crossing simple rational values.

Finally, resonances with $m=n$ associated to the $q=1$ surface are
by no way special. Sawteeth crashes could then be interpreted as ITB
collapses associated to $q=1$. A complete modeling would however
require an enlarged framework that is beyond the scope of the
present article.

\acknowledgments

Comments by A. Vasiliev on stochastic webs are gratefully
acknowledged. MCF thanks L. Nasi and J.-M. Rax for useful
discussions.

This work was carried out within the framework the European Fusion
Development Agreement and the French Research Federation for Fusion Studies.
It is supported by the European Communities under the contract of
Association between Euratom and CEA. The views and opinions expressed herein
do not necessarily reflect those of the European Commission.

\end{document}